\font\tenfrakturb=eufb10
\font\tenfraktur=eufm10
\font\tenmsbm=msbm10
\font\sevenfrakturb=eufb7
\font\sevenfraktur=eufm7
\font\sevenmsbm=msbm7
\font\fivefrakturb=eufb5
\font\fivefraktur=eufm5
\font\fivemsbm=msbm5
\newfam\bgothicfam
\newfam\gothicfam
\newfam\msbmfam
\textfont\bgothicfam = \tenfrakturb \scriptfont\bgothicfam=\sevenfrakturb
\scriptscriptfont\bgothicfam=\fivefrakturb
\textfont\gothicfam = \tenfraktur \scriptfont\gothicfam=\sevenfraktur
\scriptscriptfont\gothicfam=\fivefraktur
\textfont\msbmfam = \tenmsbm \scriptfont\msbmfam=\sevenmsbm
\scriptscriptfont\msbmfam=\fivemsbm

\def\Bbb{\tenmsbm\fam\msbmfam}

\catcode`@=11
\def\renewcounter#1{\@definecounter{#1}\@ifnextchar[{\@newctr{#1}}{}}
\documentstyle[twoside]{article}
\catcode`\@=11
\long\def\@makefntext#1{
\protect\noindent \hbox to 3.2pt {\hskip-.9pt
$^{{\eightrm\@thefnmark}}$\hfil}#1\hfill} 

\def\@makefnmark{\hbox to 0pt{$^{\@thefnmark}$\hss}} 

\def\ps@myheadings{\let\@mkboth\@gobbletwo
\def\@oddhead{\hbox{}
\rightmark\hfil\eightrm\thepage}
\def\@oddfoot{}\def\@evenhead{\eightrm\thepage\hfil
\leftmark\hbox{}}\def\@evenfoot{}
\def\sectionmark##1{}\def\subsectionmark##1{}}
\oddsidemargin=\evensidemargin
\newcounter{sectionc}\newcounter{subsectionc}\newcounter{subsubsectionc}
\renewcommand{\section}[1] {\vspace{12pt}\addtocounter{sectionc}{1}
\setcounter{subsectionc}{0}\setcounter{subsubsectionc}{0}\noindent
        {\tenbf\thesectionc. #1}\par\vspace{5pt}}
\renewcommand{\subsection}[1] {\vspace{12pt}\addtocounter{subsectionc}{1}
        \setcounter{subsubsectionc}{0}\noindent
        {\bf\thesectionc.\thesubsectionc. {\kern1pt \bfit #1}}\par\vspace{5pt}}
\renewcommand{\subsubsection}[1] {\vspace{12pt}\addtocounter{subsubsectionc}{1}
        \noindent{\tenrm\thesectionc.\thesubsectionc.\thesubsubsectionc.
        {\kern1pt \tenit #1}}\par\vspace{5pt}}
\newcommand{\nonumsection}[1] {\vspace{12pt}\noindent{\tenbf #1}
        \par\vspace{5pt}}
\newcounter{appendixc}
\newcounter{subappendixc}[appendixc]
\newcounter{subsubappendixc}[subappendixc]
\renewcommand{\thesubappendixc}{\Alph{appendixc}.\arabic{subappendixc}}
\renewcommand{\thesubsubappendixc}
        {\Alph{appendixc}.\arabic{subappendixc}.\arabic{subsubappendixc}}
\renewcommand{\appendix}[1] {\vspace{12pt}
        \refstepcounter{appendixc}
        \setcounter{figure}{0}
        \setcounter{table}{0}
        \setcounter{lemma}{0}
        \setcounter{theorem}{0}
        \setcounter{corollary}{0}
        \setcounter{definition}{0}
        \setcounter{equation}{0}
        \renewcommand{\thefigure}{\Alph{appendixc}.\arabic{figure}}
        \renewcommand{\thetable}{\Alph{appendixc}.\arabic{table}}
        \renewcommand{\theappendixc}{\Alph{appendixc}}
        \renewcommand{\thelemma}{\Alph{appendixc}.\arabic{lemma}}
        \renewcommand{\thetheorem}{\Alph{appendixc}.\arabic{theorem}}
        \renewcommand{\thedefinition}{\Alph{appendixc}.\arabic{definition}}
        \renewcommand{\thecorollary}{\Alph{appendixc}.\arabic{corollary}}
        \renewcommand{\theequation}{\Alph{appendixc}.\arabic{equation}}
        \noindent{\tenbf Appendix \theappendixc #1}\par\vspace{5pt}}
\newcommand{\subappendix}[1] {\vspace{12pt}
        \refstepcounter{subappendixc}
        \noindent{\bf Appendix \thesubappendixc. {\kern1pt \bfit #1}}
        \par\vspace{5pt}}
\newcommand{\subsubappendix}[1] {\vspace{12pt}
        \refstepcounter{subsubappendixc}
        \noindent{\rm Appendix \thesubsubappendixc. {\kern1pt \tenit #1}}
        \par\vspace{5pt}}
\newcommand{\textlineskip}{\baselineskip=13pt}
\newcommand{\smalllineskip}{\baselineskip=10pt}
\def\eightcirc{
\begin{picture}(0,0)
\put(4.4,1.8){\circle{6.5}}
\end{picture}}
\def\eightcopyright{\eightcirc\kern2.7pt\hbox{\eightrm c}}
\newcommand{\copyrightheading}[1]
        {\vspace*{-2.5cm}\smalllineskip{\flushleft
        {\footnotesize Modern Physics Letters A, #1}\\
        {\footnotesize $\eightcopyright$\, World Scientific Publishing
         Company}\\
         }}
\newcommand{\pub}[1]{{\begin{center}\footnotesize\smalllineskip
        Received #1\\
        \end{center}
        }}

\def\abstracts#1#2#3{{
        \centering{\begin{minipage}{4.5in}\baselineskip=10pt\footnotesize
        \parindent=0pt #1\par
        \parindent=15pt #2\par
        \parindent=15pt #3
        \end{minipage}}\par}}

\newcommand{\bibit}{\nineit}
\newcommand{\bibbf}{\ninebf}
\renewenvironment{thebibliography}[1]
         {\frenchspacing
         \ninerm\baselineskip=11pt
         \begin{list}{\arabic{enumi}.}
         {\usecounter{enumi}\setlength{\parsep}{0pt}
         \setlength{\leftmargin 12.7pt}{\rightmargin 0pt} 
         \setlength{\itemsep}{0pt} \settowidth
         {\labelwidth}{#1.}\sloppy}}{\end{list}}
\newcounter{itemlistc}
\newcounter{romanlistc}
\newcounter{alphlistc}
\newcounter{arabiclistc}

\newcommand{\fcaption}[1]{
         \refstepcounter{figure}
         \setbox\@tempboxa = \hbox{\footnotesize Fig.~\thefigure. #1}
         \ifdim \wd\@tempboxa > 5in
           {\begin{center}
         \parbox{5in}{\footnotesize\smalllineskip Fig.~\thefigure. #1}
            \end{center}}
        \else
             {\begin{center}
             {\footnotesize Fig.~\thefigure. #1}
              \end{center}}
        \fi}
\newcommand{\tcaption}[1]{
        \refstepcounter{table}
        \setbox\@tempboxa = \hbox{\footnotesize Table~\thetable. #1}
        \ifdim \wd\@tempboxa > 5in
           {\begin{center}
        \parbox{5in}{\footnotesize\smalllineskip Table~\thetable. #1}
            \end{center}}
        \else
             {\begin{center}
             {\footnotesize Table~\thetable. #1}
              \end{center}}
        \fi}
\def\@citex[#1]#2{\if@filesw\immediate\write\@auxout
        {\string\citation{#2}}\fi
\def\@citea{}\@cite{\@for\@citeb:=#2\do
        {\@citea\def\@citea{,}\@ifundefined
        {b@\@citeb}{{\bf ?}\@warning
        {Citation `\@citeb' on page \thepage \space undefined}}
        {\csname b@\@citeb\endcsname}}}{#1}}
\newif\if@cghi
\def\cite{\@cghitrue\@ifnextchar [{\@tempswatrue
        \@citex}{\@tempswafalse\@citex[]}}
\def\citelow{\@cghifalse\@ifnextchar [{\@tempswatrue
        \@citex}{\@tempswafalse\@citex[]}}
\def\@cite#1#2{{$\null^{#1}$\if@tempswa\typeout
        {IJCGA warning: optional citation argument
        ignored: `#2'} \fi}}

\def\pmb#1{\setbox0=\hbox{#1}
        \kern-.025em\copy0\kern-\wd0
        \kern.05em\copy0\kern-\wd0
        \kern-.025em\raise.0433em\box0}

\def\fnt#1#2{\footnotetext{\kern-.3em
        {$^{\mbox{\scriptsize #1}}$}{#2}}}
\def\fpage#1{\begingroup
\voffset=.3in
\thispagestyle{empty}\begin{table}[b]\centerline{\footnotesize #1}
       \end{table}\endgroup}
\def\runninghead#1#2{\pagestyle{myheadings}
\markboth{{\protect\footnotesize\it{\quad #1}}\hfill}
{\hfill{\protect\footnotesize\it{#2\quad}}}}
\font\tenrm=cmr10
\font\tenit=cmti10
\font\tenbf=cmbx10
\font\bfit=cmbxti10 at 10pt
\font\ninerm=cmr9
\font\nineit=cmti9
\font\ninebf=cmbx9
\font\eightrm=cmr8

\def\qed{\hbox{${\vcenter{\vbox{  
   \hrule height 0.4pt\hbox{\vrule width 0.4pt height 6pt
   \kern5pt\vrule width 0.4pt}\hrule height 0.4pt}}}$}}

\begin{document}
\def\bh{${\Bbb R}^2\times {\Bbb S}^2\>$}
\def\ghc{{\sqrt G\over\hbar c}}
\runninghead{N. E. Firsova}
{S-matrices for  spinor particles
 on Reissner-Nordstr\"{o}m black holes}
\normalsize\textlineskip
\thispagestyle{empty}
\setcounter{page}{1}
\copyrightheading{Vol. 18, No. 19 (2003) 1287-1296}
\vspace*{0.88truein}
\fpage{1}
\centerline{\bf S-MATRICES FOR SPINOR PARTICLES}
\vspace*{0.035truein}
\centerline{\bf ON REISSNER-NORDSTR\"{O}M BLACK HOLES}
\vspace*{0.035truein}
\vspace*{0.37truein}
\centerline{\footnotesize N. E. FIRSOVA}
\vspace*{0.015truein}
\centerline{\footnotesize\it Institute for Mechanical Engineering,
Russian Academy of Sciences}
\baselineskip=10pt
\centerline{\footnotesize\it Sankt-Petersburg 199178, Russia}
\vspace*{0.225truein}
\pub{January 2002}
\vspace*{0.21truein}
\abstracts{ The scattering problems arising
when considering the contribution of
the topologically inequivalent configurations of the 
spinors on Reissner-Nordstr\"{o}m  black holes to the Hawking radiation
are correctly stated. The corresponding $S$-matrices are described and
presented in the form convenient to numerical computations.
}{}{}
\vspace*{1pt}\textlineskip 
\section{Introduction} 
\vspace*{-0.5pt}
\noindent
The nontrivial topological properties of black holes
may play essential role while studying quantum geometry of fields on them.
Really the black holes can
carry the whole spectrum of topologically inequivalent configurations
(TICs)
for miscellaneous fields, in the first turn, complex scalar and spinor
ones.
The mentioned TICs can markedly modify the Hawking radiation from black
holes. 

Physically, the existence of TICs should
be obliged to the natural presence of magnetic U(N)-monopoles
(with $N\ge1$)
on black holes though the total (internal) magnetic charge (abelian or
nonabelian) of black hole remains equal to zero.
One can consider that monopoles reside in black holes
as quantum objects without having influence on the black hole metrics.
They can reside in the form of monopole gas in which the process
of permanent creation and annihilation of the virtual monopole-antimonopole
pairs occurs so that the summed internal magnetic charge (i.e., connected with
topological properties) is equal to zero while the external (not
connected with topological properties) one may differ from zero.

While existing the virtual monopole-antimonopole pair can interact
with a particle and, by this, increasing the Hawking radiation.

This influence on the Hawking radiation has been studied more or less
for the TICs of complex scalar field (see Ref.\cite{GF} and
references therein for more details). 

Investigation of the spinor
case has only recently started. A description of TICs for
spinors was obtained in Ref.\cite{Gon99}. The detailed analysis of
their contribution to Hawking radiation required the knowledge of the
conforming $S$-matrices which regulate the spinor particle passing through the
potential barrier surrounding black hole.
Those $S$-matrices were explored in Ref.\cite{Fir2001} for the case
of  spinors on Schwarzschild black holes.
An algorithm to calculate
their elements numerically was gained and applied to the calculation
of all the configurations luminosity for massless spinors from Schwarzschild
black holes.\cite{GF01}

The present paper is just devoted to building the mentioned $S$-matrices
for the  spinor TICs on Reissner-Nordstr\"{o}m (RN) black holes
and obtaining the
appropriate algorithm to numerically compute their elements.

The metric under consideration can be written down in the form
$$ds^2=g_{\mu\nu}dx^\mu\otimes dx^\nu\equiv
adt^2-a^{-1}dr^2-r^2(d\vartheta^2+\sin^2\vartheta d\varphi^2) \eqno(1)$$
with $a=1-2M/r+\alpha^2M^2/r^2$, $\alpha=Q/M$, where $M$, $Q$ are respectivly
a black hole mass and a charge. Besides we have $|g|=|det(g_{\mu\nu})|=
(r^2\sin \vartheta)^2$, $r_{\pm}=M(1\pm\sqrt{1-\alpha^2})$ with
$0\leq\alpha\leq 1$, so $r_+\leq r<\infty$, $0\leq\vartheta<\pi$,
$0\leq \varphi<2\pi$.

  Throughout the paper we employ the system of units with $\hbar=c=G=1$,
unless explicitly stated otherwise. Our choice of the Dirac
$\gamma$-matrices is the same as in Ref.\cite{Gon99}
Finally, we shall denote $L_2(F)$ the set of the modulo square integrable
complex functions on any manifold $F$ furnished with an integration measure
while $L^n_2(F)$ will be the $n$-fold direct product of $L_2(F)$
endowed with the obvious scalar product.

\section{Preliminary remarks}
  As was discussed in Ref.\cite{Gon99}, TICs of spinor field on the
black holes under consideration
arise due to existence of the twisted spinor bundles
over the standard black hole topology \bh. From a physical point of view
the appearance of spinor twisted configurations is linked with the natural
presence of Dirac monopoles that play the role of connections in the
complex
line bundles corresponding to the twisted spinor bundles.
Under this situation, each such a bundle can be labeled
by the Chern number $n\in{\Bbb Z}$ and
the wave equation for corresponding massive spinors $\Psi$ as
sections of the mentioned bundle looks as follows
$${\cal D}_n\Psi=\mu_0\Psi,\>\eqno(2)$$
with the twisted Dirac operator ${\cal D}_n=i\gamma^\mu\nabla_\mu^n$
and we can call (standard) spinors corresponding to $n=0$
{\it untwisted} while the rest of the spinors with $n\ne0$
should be referred to as {\it twisted}. Referring for details and for
explicit form of ${\cal D}_n$ to Ref.\cite{Gon99}, it should be noted
here
that in $L_2^4$(\bh) there is a basis from the solutions of (2) in the form
$$\Psi_{\lambda m}=\frac{1}{\sqrt{2\pi\omega}}
e^{i\omega t}r^{-1}\pmatrix{F_1(r,\omega,\lambda)
\Phi_{\lambda m}\cr
F_2(r,\omega,\lambda)\sigma_1\Phi_{\lambda m}\cr}\>, \eqno(3)$$
where $\sigma_1$ is the Pauli matrix, the 2D spinor
$\Phi_{\lambda m}=\Phi_{\lambda m}(\vartheta,\varphi)=
(\Phi_{1\lambda m},\Phi_{2\lambda m})$ is
the eigenspinor
of the twisted euclidean Dirac operator with Chern number $n$ on the unit
sphere with the eigenvalue $\lambda =\pm\sqrt{(l+1)^2-n^2}$ while
$-l\le m\le l+1$, $l\ge|n|$. The functions $F_{1,2}$ obey the system of
equations
$$\cases{\sqrt{a}\partial_rF_1+
\left(\frac{1}{2}\frac{d\sqrt{a}}{dr}+\frac{\lambda}{r}\right)F_1=
-i[\mu_0-\frac{\omega+e_0 Q/r}{\sqrt{a}}]F_2,\cr
\sqrt{a}\partial_rF_2+
\left(\frac{1}{2}\frac{d\sqrt{a}}{dr}-\frac{\lambda}{r}\right)F_2=
-i[\mu_0+\frac{\omega+e_0 Q/r}{\sqrt{a}}]F_1\cr} \>\eqno(4)$$
The explicit form of the 2D spinor
$\Phi_{\lambda m}$ is inessential in the given paper and can be found in
Ref.\cite{Gon99}. One can only notice here that
they can be subject to the normalization condition at $n$ fixed
$$\int\limits_0^\pi\,\int\limits_0^{2\pi}(|\Phi_{1\lambda m}|^2+
|\Phi_{2\lambda m}|^2)
\sin\vartheta d\vartheta d\varphi=1$$
and these spinors form an orthonormal basis in $L_2^2({\Bbb S}^2)$ at any
$n\in{\Bbb Z}$.

By passing on to the variable
$$r_*=r+\frac{r_+^2}{r_+-r_-}\ln|\frac{r-r_+}{2M}|-
\frac{r_-^2}{r_+-r_-}\ln|\frac{r-r_-}{2M}|$$
where $r_{\pm}=M\pm\sqrt{M^2-Q^2}$ and by
going to the quantities $x=r_*/M, y=r/M, k=\omega M,\mu=\mu_0 M,
y_{\pm}=r_{\pm}/M=1\pm\sqrt{1-\alpha^2}$, we shall have
$$x=y+\frac{y_+^2}{y_+-y_-}\ln|\frac{y-y_+}{2M}|-
\frac{y_-^2}{y_+-y_-}\ln|\frac{y-y_-}{2M}|$$
so that $y(x)$ is given implicitly by the latter relation and
$$y'=dy/dx=1-2/y+\alpha^2/r^2=(y-y_+)(y-y_-)/y^2=a=a(x,\alpha)\>\eqno(5)$$
and the system (4) can be rewriten as follows
$$\cases{E'_1 +a_1 E_1=b_1 E_2\>,\cr
 E'_2 + a_2 E_2=b_2E_1\cr} \>\eqno(6)$$
with $E_{1}=E_{1}(x,k,\lambda)=F_+(Mx)$,
$F_+(r^*)=F_{1}[r(r^*)]$,
$E_{2}=E_{2}(x,k,\lambda)=iF_-(Mx)$,
$F_-(r^*)=F_{2}[r(r^*)]$
and
$$a_{1,2}=\frac{1}{2y^2}(1-\frac{\alpha^2}{y})
\pm\frac{\lambda}{y}\sqrt{a}\>,\eqno(7) $$
$$b_{1,2}=\mu\sqrt{a}\mp(k+\frac{e\alpha}{y}),
\qquad e=e_0 M \>.\eqno(8) $$

Let us eliminate $E_2$ to find equation for $E_1$. Without going into
details we obtain
$$E''_1+\tilde{a}E_1'+\tilde{b}E_1=(\mu^2-k^2) E_1\>,\eqno(9)$$
where
$$\tilde{a}=\frac{1}{y^2}(1-\frac{\alpha^2}{y})-\frac{b'_1}{b_1}\>,\eqno(10)$$
$$\tilde{b}=2k\frac{e\alpha}{y}+
\frac{e^2\alpha^2-\lambda ^2 a}{y^2}+
\frac{1}{4y^4}(1-\frac{\alpha^2}{y})^2+a'_1-\frac{b'_1}{b_1}a_1\>.\eqno(11)$$
Now we apply the ansatz
$$E_1=u_1\exp\left[-\frac{1}{2}\int \tilde{a}dx\right] \>\eqno(12)$$
and obtain instead of (9)
the following Schr\"odinger-like equation for the function $u_1$
$$u''_1+(k^2-\mu^2)u_1=q_1u_1 \>\eqno(13)$$
where
$$q_1=\frac{1}{4}\tilde{a}^2+\frac{1}{2}\tilde{a}'-\tilde{b}\>.\eqno(14)$$
With the help of the relation
$$\int\frac{1}{y^2}(1-\frac{\alpha^2}{y})dx=
\int \frac{(y-\alpha^2)dy}{y(y-y_+)(y-y_-)}$$
we transform the (12) as follows
$$E_1=iu_1\sqrt{\frac{k+e\alpha/y}{\sqrt{a}}-\tilde{\mu}}\>.
\eqno(15)$$
The (14) can be rewritten in the form
$$q_1=-2\frac{\mu^2+
e\alpha k}{y}+\frac{(\mu^2-e^2)\alpha^2}{y^2}+
\frac{a\lambda^2 }{y^2}+\frac{a(3\alpha^2-2y)}{2y^4}+$$
$$\frac{\lambda\sqrt{a}}{y}\frac{b_1'}{b_1}+
\frac{3}{4}({\frac{b_1'}{b_1}})^2-\frac{1}{2}\frac{b_1''}{b_1}-a_1'\>.
\eqno(16)$$
Consequently we can transform (13) to the following
$$u''_1+[(k+\frac{e\alpha}{y})^2-a\mu^2]u_1=
\frac{\sqrt{a}\lambda^2}{y^2}\tilde{q}_1u_1\>.\eqno(17)$$
Here
$$\tilde{q}_1=\sqrt{a}+\frac{\sqrt{a}(3\alpha^2-2y)}{2y^2\lambda^2}+
\frac{y}{\lambda}\frac{b'_1}{b_1}+$$
$$\frac{3}{4}\frac{y^2}{\lambda^2\sqrt{a}}(\frac{b'_1}{b_1})^2-
\frac{1}{2}\frac{y^2}{\lambda^2\sqrt{a}}\frac{b''_1}{b_1}-
\frac{y^2}{\lambda^2\sqrt{a}}a'_1\>,\eqno(18)$$
where
$$\frac{b'_1}{b_1}=-\frac{\sqrt{a}}{y^2}
\frac{\mu(1-\frac{\alpha^2}{y})+e\alpha \sqrt{a}}
{k+\frac{e\alpha}{y}-\mu\sqrt{a}}$$
$$\frac{b_1''}{b_1}=
-\frac{\sqrt{a}}{y^3}
\frac{\mu(-2+\frac{5+3\alpha^2}{y}-\frac{10\alpha^2}{y^2}+
\frac{4\alpha^4}{y^3})-
2e\alpha(1-\frac{3}{y}+\frac{2\alpha^2}{y^2})}
{k+\frac{e\alpha}{y}-\mu\sqrt{a}}$$
$$a'_1=-\frac{\lambda\sqrt{a}}{y^2}
\left[1-\frac{3}{y}+\frac{2\alpha^2}{y^2}+
\frac{\sqrt{a}}{\lambda y}(1-\frac{3\alpha^2}{2y})\right]$$
So it is not difficult to see that the function $\tilde{q}_1$ is bounded
uniformly in
$x\in (-\infty,\infty)$, $k\geq \mu$ when $\lambda\to\infty$, i.e.
$$|\tilde{q}_1|\leq C.\>\eqno(19)$$

\section{The scattering problem}
One can notice that though the potentials $q_1=q_1(x,k,\lambda)$ and
$\tilde{q}_1=\tilde{q}_1(x,k,\lambda)$
of (16),(18) is given implicitly, it satisfies
the conditions explored in Ref.\cite{Fir99} and referring for more details
to those works we may here use their results to formulate correctly the
scattering problem for (17). Namely, the correct statement of the
scattering problem will consist in searching for two
solutions $u_1^+(x,k,\lambda)$,
$u_1^-(x,k,\lambda)$ of the equation (17)
obeying the following conditions

$$u_1^+(x,k,\lambda)=
\cases{e^{i(k+e\alpha/y_+)x}+
s_{12}(k,\lambda)e^{-i(k+e\alpha/y_+)x}+o(1),&$x\to-\infty$,\cr
s_{11}(k,\lambda)w_{i\beta,\frac{1}{2}}(-2ik^+x)+
o(1),&$x\to+\infty$,\cr}$$

$$u_1^-(x,k,\lambda)=\cases{s_{22}(k,\lambda)e^{-i(k+e\alpha/y_+)x}
+o(1),&$x\to-\infty$,\cr
w_{-i\beta,\frac{1}{2}}(2ik^+x)+
s_{21}(k,\lambda)w_{i\beta,\frac{1}{2}}(-2ik^+x)+
o(1),&$x\to+\infty$,\cr}
\eqno(20)$$
where
$$\beta=\frac{\mu^2+e\alpha k}{k^+}\qquad
k^+=\sqrt{k^2-\mu^2}$$
and the functions  $w_{\pm i\beta,\frac{1}{2}}(\pm z)$
are related to the Whittaker functions
$W_{\pm i\beta,\frac{1}{2}}(\pm z)$
(concerning the latter ones see e. g. Ref.\cite{Abr64}) by the relation
$$w_{\pm i\beta,\frac{1}{2}}(\pm z)=
W_{\pm i\beta,\frac{1}{2}}(\pm z)e^{-\pi\beta/2}\>,$$
so that one can easily gain asymptotics (using the corresponding ones for
Whittaker functions \cite{Abr64})

$$w_{i\beta,\frac{1}{2}}(-2ik^+x)=
e^{ik^+x}e^{i\beta\ln|2k^+x|}[1+O(|k^+x|^{-1})], \> x\to+\infty\>,$$
$$w_{-i\beta,\frac{1}{2}}(2ik^+x)=e^{-ik^+x}
e^{-i\beta,\ln|2k^+x|}[1+O(|k^+x|^{-1})], \> x\to+\infty\>.\eqno(21)$$

We can see that there arises some $S$-matrix with elements
$s_{ij}, i, j = 1, 2$.
As will be seen below, for calculating the Hawking
radiation we need the coefficient $s_{11}$, consequently, we need to have
some algorithm for numerical computation of it inasmuch as the latter
cannot
be evaluated in exact form. The given algorithm can be extracted from the
results of Ref.\cite{Fir99}. To be more precise
$$s_{11}(k,\lambda)=2i\left(k+\frac{e\alpha}{y_+}\right)
/[f^-(x,k,\lambda),f^+(x,k^+,\lambda)]\>,\eqno(22)$$
where [,] signifies the Wronskian of functions $f^-,f^+$, the
so called Jost type solutions of (17).
In their turn, these functions and
their derivatives obey the certain integral equations. Since the Wronskian
does not depend on $x$ one can take the mentioned
integral equations by $x=x_0$

$$f^-(x,k,\lambda)=e^{-i(k+e\alpha /y_+)x}+$$
$$\frac{1}{k+e\alpha /y_+}\int\limits^{x}_{-\infty}
\sin[(k+e\alpha/y_+)(x-t)]q_1^-(t,k,\lambda)f^-(t,k,\lambda)dt\>,
\eqno(23)$$
$$(f^-)'_x(x,k,\lambda)=
-i(k+e\alpha /y_+)e^{-i(k+e\alpha/y_+)x}+$$
$$\int\limits^{x}_{-\infty}\cos[(k+e\alpha /y_+)
(x-t)]q_1^-(t,k,\lambda)f^-(t,k,\lambda)dt\>,\eqno(24)$$
and
$$f^+(x,k^+,\lambda)=w_{i\beta,\frac{1}{2}}(-2ik^+x)+$$
$$\frac{1}{k^+}
\int\limits_{x}^{+\infty}{\rm Im}[w_{i\beta,\frac{1}{2}}(-2ik^+x)
w_{-i\beta,\frac{1}{2}}(2ik^+t)]
q_1^+(t,k,\lambda)f^+(t,k^+,\lambda)dt\>,\eqno(25)$$
$$(f^+)'_x(x,k^+,\lambda)=
\frac{d}{dx}w_{i\beta,\frac{1}{2}}(-2ik^+x)+$$
$$\frac{1}{k^+}\int\limits^{+\infty}_{x}{\rm Im}\left[\frac{d}{dx}
w_{i\beta,\frac{1}{2}}(-2ik^+x)
w_{-i\beta,\frac{1}{2}}(2ik^+t)\right]q_1^+(t,k,\lambda)
f^+(t,k^+,\lambda)dt\>,\eqno(26)$$
where the potentials have the form
$$q_1^-(x,k,\lambda)=\frac{\lambda^2\sqrt{a}}{y^2}\tilde{q}_1+
a\left(\mu^2+\frac{e\alpha}{yy_+(y-y_+)}
\left(2k+e\alpha(y^{-1}+y_+^{-1})\right)\right)
\>\eqno(27)$$
and
$$q_1^+(x,k,\lambda)=\frac{\lambda^2\sqrt{a}}{y^2}\tilde{q}_1+
2(\mu^2+e\alpha k)\frac{y-x}{xy}+
\frac{\alpha^2(\mu^2-e^2)}{y^2}
\>\eqno(28)$$
The potential $q_1^-$ exponencially tends to zero when
$x\to-\infty$ and $q_1^+$ behaves as $O(x^{-2})$ as
$x\to+\infty$. So one can notice that these potentials
are integrable when $x\to-\infty$ or $x\to\infty$ respectively.
The point $x=x_0$ should be chosen from the considerations of
the computational convenience.
The relations (22)--(28) can be employed for numerical calculation of
$s_{11}$ (see Ref.\cite{GF} for the complex scalar field case).

To summarize, we can now obtain
the general solution of (17) in the class of
functions restricted on the whole $x$-axis in the linear combination form
$$u_1(x,k,\lambda)=C^1_{\lambda}(k)u_1^+(x,k,\lambda)+
C^2_{\lambda}(k)u_1^-(x,k,\lambda)\>,\eqno(29)$$
since a couple of the functions $u_1^+(x,k,\lambda)$,
$u_1^-(x,k,\lambda)$
forms a fundamental system of solutions for the equation (17).
Besides we have the unitarity relations
(for more details see Ref.\cite{Fir99})
$$\frac{1}{\gamma(k)}|s_{22}(k,\lambda)|^2+|s_{21}(k,\lambda)|^2=1,\qquad
 \gamma(k)|s_{11}(k,\lambda)|^2+|s_{12}(k,\lambda)|^2=1,\qquad
\gamma(k)=\frac{k^+}{k}$$
and it is not difficult to show that the following asymptotic behavior
occurs
$$s_{12}(k,\lambda)=O(k^{-1}),\qquad s_{21}(k,\lambda)=O(k^{-1})\>,$$
$$\Gamma(k,\lambda)=|s_{11}(k,\lambda)|^2=1+O(k^{-1}).$$
When analysing the Hawking radiation one should put $C^2_{\lambda}(k)=0$
and then using (20)--(21) we can find asymptotic of the function $E_1$
as $x\to\infty$
$$E_1=C^1_\lambda(k)i\sqrt{k-\mu}s_{11}(k,\lambda)
e^{ik^+x}e^{i\beta\ln(2k^+x)}+o(1),\qquad x\to\infty \>.\eqno(30)$$

To study the Hawking radiation we shall also need asymptotic of the
function $E_2$ at $x\to\infty$.
To find the latter function we should use the first
equation of (6) so that
$$E_2=\frac{1}{b_1}(E'_1+a_1E_1)$$
and after differentiating the relation (15) we obtain
$$E_2=C^1_{\lambda}(k)\sqrt{k+\mu}s_{11}(k,\lambda)
e^{ik^+x}e^{i\beta\ln(2k^+x)}+o(1),
\qquad x\to\infty \>.\eqno(31)$$
At last, with the help of (30) and (31) it is easily to establish that
at $x\to+\infty$
$$E_1^*\partial_x E_1+E_2^*\partial_xE_2=
-(E_1\partial_x E_1^*+E_2\partial_xE_2^*)=
-2ik^+|s_{11}(k,\lambda)|^2|C^1_\lambda(k)|^2+o(1)\>,$$
$$E_1^*E_2-E_2^*E_1=-2ik^+|s_{11(k,\lambda}|^2|C^1_\lambda(k)|^2+o(1)
\>,\eqno(32)$$
where (*) means complex conjugation.

 Though for studying the Hawking radiation one needs only $s_{11}$, the
other elements of the $S$-matrix described can be important in a number of
the problems within the 4D black hole physics, for instance, when studying
vacuum polarization near black holes for spinor TICs.

Having obtained all the above, we can discuss the Hawking radiation
process for any TIC of spinor field.

\section{Modification of Hawking radiation}
  As was pointed out in Ref.\cite{Gon99}, the equation (2) corresponds
to the lagrangian
$${\cal L}=
\frac{i}{2}|g|^{1/2}\left[\overline{\Psi}\gamma^\mu\nabla_\mu^n\Psi-
(\nabla_\mu^n\overline{\Psi})\gamma^\mu\Psi\right]
\>, \eqno(33)$$
where $\overline{\Psi}=\Psi^{\dag}\gamma_0$ is the adjont spinor and
($\dag$)
stands for hermitian conjugation.
As a result, we have the conforming energy-momentum tensor
for TIC with the Chern number $n$ conforming to the lagrangian (34)
$$T_{\mu\nu}(\overline{\Psi},\Psi)=
\frac{i}{4}\left[\overline{\Psi}\gamma_\mu\nabla_\nu^n\Psi+
\overline{\Psi}\gamma_\nu\nabla^n_\mu\Psi-
(\nabla_\mu^n\overline{\Psi})\gamma_\nu\Psi-
(\nabla_\nu^n\overline{\Psi})\gamma_\mu\Psi\right] \>.\eqno(34) $$

When quantizing twisted TIC with the Chern number $n$ we shall
take the set of spinors
$\Psi_{\lambda m}$ of (3) as a basis in $L_2^4$(\bh) so that
$-l\le m\leq l+1$, $l=|n|,|n|+1$,... and we can evidently realize
the procedure of quantizing
TIC with the Chern number $n$, as usual, by expanding in the modes
$\Psi_{\lambda m}$
$$\Psi=\sum\limits_{\pm\lambda}\sum\limits_{l=|n|}^\infty
\sum\limits_{m=-l}^{l+1}
\int\limits_{\mu}^\infty\,d\omega
(a^-_{\omega nlm}\Psi_{\lambda m}+
b^+_{\omega nlm}{\Psi_{-\lambda m}})\,,$$
$$\overline{\Psi}=
\sum\limits_{\pm\lambda}\sum\limits_{l=|n|}^\infty\sum\limits_{m=-l}^{l+1}
\int\limits_{\mu}^\infty\,d\omega
(a^+_{\omega nlm}\overline{\Psi}_{\lambda m}+
b^-_{\omega nlm}\overline{\Psi}_{-\lambda m})\,,\eqno(35)$$
so that $a^\pm_{\omega nlm}$, $b^\pm_{\omega nlm}$ should be interpreted
as the corresponding
creation and annihilation operators for spinor particle in both the
gravitational field of black hole and the field of the conforming monopole
with the Chern number $n$ and we have the standard anticommutation
relations
$$\{a^-_i,\,a^+_j\}=\delta_{ij},\>\{b^-_i,\,b^+_j\}=\delta_{ij}\>
\eqno(36)$$
and zero for all other anticommutators,
where $i =\{\omega nlm\}$ is a generalized index.

  Under the circumstances one can speak about the Hawking radiation
process for any TIC of spinor field.
To get the luminosity $L(n)$ with respect to the Hawking radiation for TIC
with the Chern number $n$, we define a vacuum state $|0>$ by the conditions
$$a^-_i|0>=0, \> b^-_i|0>=0\>,\eqno(37)$$
and then

$$L(n)=\lim_{r\to\infty}\,\int\limits_{S^2}\,
<0|T_{tr}|0>d\sigma \eqno(38)$$
with the vacuum expectation value $<0|T_{tr}|0>$ and the surface element
$d\sigma=r^2\sin\vartheta d\vartheta\wedge d\varphi$. Now with the help
of (35)--(38) we have
$$L(n)=\lim_{r\to\infty}\,\int\limits_{S^2}\,
\left[\sum\limits_{\pm\lambda}\sum\limits_{l=|n|}^\infty
\sum\limits_{m=-l}^{l+1}
\int\limits_{\mu}^\infty\,T_{tr}(\overline{\Psi_{\lambda m}},
\Psi_{\lambda m})d\omega\right]d\sigma \eqno(39)$$
with spinors $\Psi_{\lambda m}$ of (3). After this we can fix vacuum state
by choosing
$$C^1_\lambda(k)=\sqrt{\frac{B(\alpha)}
{e^{4\pi k[1+f(\alpha)]}+B(\alpha)}}\>$$
with
$f(\alpha)=2^{-1}(\sqrt{1-\alpha^2}+1/\sqrt{1-\alpha^2})$,
$B(\alpha)=
\exp[-4\pi e\alpha M\ghc\frac{1+f(\alpha)}{1+\sqrt{1-\alpha^2}}]$,
$e=4.8\cdot10^{-10}\,{\rm cm^{3/2}\cdot g^{1/2}\cdot s^{-1}}$,
and we shall find (in usual units)
$$L(n)=
A\sum\limits_{\pm\lambda}\sum\limits_{l=|n|}^\infty2(l+1)B(\alpha)
\int\limits_{\mu}^{\infty}\frac{|s_{11}(k,\lambda)|^2}
{e^{4\pi k[1+f(\alpha)]}+B(\alpha)}k^+dk, \>\eqno(40)$$
where the relations
$$\frac{d}{dr}=\frac{1}{a}\frac{d}{dr^*}=\frac{1}{aM}
\frac{d}{dx},\>$$
$$ \gamma_t=\sqrt{a}\gamma^0,\>$$
$$\gamma_r=-\frac{1}{\sqrt{a}}\gamma^1  $$
have been used while
$A=\frac{c^5}{\pi GM}\left(\frac{c\hbar}{G}\right)^{1/2}
\approx0.251455\cdot10^{55}\,
{\rm{erg\cdot s^{-1}}}\cdot M^{-1}$,
$M$ in g.
It should be
noted that in (41), generally speaking,
$s_{11}(k,\lambda)\ne s_{11}(k,-\lambda)$ and that is clear from the form
of potential $q_1$ of (17).

We can interpret $L(n)$, as usual,\cite{{GF},{Gon99}} as an additional
contribution to the Hawking radiation due to the additional spinor
particles leaving the black hole because of the interaction with
monopoles and the conforming radiation
can be called {\it the monopole Hawking radiation}\cite{Gon99}.

Under this situation,
for the all configurations luminosity $L$ of black hole with respect
to the Hawking radiation concerning the spinor field to be obtained,
one should sum up over all $n$, i. e.
$$L=\sum\limits_{n\in{\Bbb{Z}}}\,L(n)=L(0)+
2\sum\limits_{n=1}^\infty\,L(n)
\eqno(41)$$
since $L(-n)= L(n)$. The convergence of the series of (41)--(42),
respectively, over $l$ and $n$ can be proved by the same considerations
as in Ref.\cite{GF} what we shall here not dwell upon.

As a result, we can expect a marked increase of Hawking radiation from
black holes under consideration for spinor particles.
But for to get an exact value of this increase one should
apply numerical methods.

\section{Conclusion}
The relations obtained here can be employed when numerically
calculating the Hawking radiation luminosity for the really existing spinor
particles, for example, for $e^\pm$ and neutrinos and also for other
fundamental leptons.

In Ref.\cite{GF01} it was calculated in the Schwarzschild black hole case
that the additional Hawking radiation due to twisted massless spinors is about
22 per cent. It would be interesting to have known as this result will change
for Reissner--Nordstr{\"o}m black hole in dependence of its electric charge
$Q$. We hope to discuss the numerical results elsewhere.

\nonumsection{Acknowledgement}
\noindent
     The author is thankful to Yu. Goncharov for critical reading of the 
manuscript.

\newpage
\nonumsection{References}

\end{document}